\documentstyle[aps,multicol,epsfig]{revtex}

\begin{document}

\title{Implementation of Conditional Phase Shift gate for Quantum Information Processing by NMR,
        using Transition-selective pulses.}

\author{Ranabir Das$^\dagger$, T.S. Mahesh $^\dagger$ and Anil Kumar $^{\dagger,\ddagger}$\\
        $^{\dagger}$ {\it Department of Physics,}
        $^{\ddagger}$ {\it Sophisticated Instruments Facility}\\
        {\it Indian Institute of Science, Bangalore 560012 India}\\}

\maketitle

\begin{abstract}
 Experimental realization of quantum information processing in the field 
of nuclear magnetic resonance (NMR) has been well established. 
Implementation of conditional phase shift gate has been 
a significant step, which has lead to realization of important algorithms such as 
Grover's search algorithm and quantum Fourier transform. This gate has so far been 
implemented in NMR by using coupling evolution method. We demonstrate here the 
implementation of the conditional phase shift gate using transition selective pulses.
 As an application of the gate, we demonstrate Grover's search algorithm and quantum Fourier 
transform  by simulations and experiments using transition selective pulses.      
\end{abstract}

\begin{multicols}{2}
\section{Introduction}
Theoretical possibility of quantum information processing (QIP) has 
generated a lot of enthusiasm for its experimental realization 
\cite{rf,deu,pw,ss,gr,pre,db,ic}. 
Nuclear magnetic resonance (NMR) has played a leading role for practical demonstration 
of quantum algorithms and quantum gates [10-32]. Deutsch-Jozsa algorithm\cite{na,dg,kd,lo},
Grover's search algorithm\cite{ci,jo,ap}, quantum Fourier transform\cite{qft}
 and the Shor's factorization algorithm\cite{nat} have been implemented by liquid-state 
NMR. The unitary operators needed for implementation of these algorithms by NMR, have 
mostly been realized using spin selective radio frequency (r.f) 
pulses and coupling evolution, utilizing indirect spin-spin(J) or 
dipolar couplings among the spins. On the other hand several logic gates and algorithms have also been implemented 
using transition selective (soft; low power, long duration) r.f pulses \cite{dg,kd,ne,ak,ka}. 
 The use of transition selective pulses in 
quantum information processing is popular for its simplicity of logical operations. For 
example, a C$^2$-NOT gate in a three qubit system using coupling evolution, 
requires a series of spin (qubit) selective  
$\pi/2$ and $\pi$ pulses inter-spaced with J-evolutions on all three qubits, cascading a series of 
unitary transforms, while the same gate needs a single 
transition selective  $\pi$ pulse\cite{dg}. 
 
 Conditional phase shift gate is an integral part of many algorithms such as Grover's 
search algorithm and quantum Fourier transform(QFT)\cite{pre}. This gate introduces a phase-shift only if 
a certain condition is fulfilled. This gate has been  realized using J-evolution by 
earlier workers \cite{ci,ap,qft}. In this work, we construct conditional phase shift gate 
using transition selective pulses. As an application we demonstrate Grover's search algorithm 
and quantum Fourier transform in a 3 qubit system by simulations, and  Grover's 
search algorithm in a 2 qubit system by experiments, using transition selective pulses. 

 It may be mentioned that while theoritically the transition selective pulses are attractive 
since they simplify the logic of an operation, their experimental implementation requires long 
 low power r.f. pulses which give rise to experimental errors due to relaxation and unwanted 
evolution under the internal Hamiltonian during long pulses \cite{pou}. However, the  
transition selective pulse method has yielded experimental results with fidelity comparable to 
J-evolution method \cite{ts}. Recently, use of transition selective pulses for quadrupolar nuclei of spin 3/2 and 7/2 
 partially oriented in liquid crystal matrices respectively as 2 and 3 qubit systems has been demonstrated \cite{fu1,ne,mur}. 
 Transition selective pulses have also been used to demonstrate the use of 
 oriented dipolar coupled CH$_3$ and $^{13}$CH$_3$ groups as 2 and 3 qubits  respectively and 
work is in progress to use transition selective pulses for quantum information processing 
in strongly coupled spins \cite{fu,ak,mah}. In all these cases the J-evolution method is either not applicable 
or too complex to implement \cite{abab}. In such situations the use of transition selective pulses method,
 inspite of its experimental limitations, provides an attractive alternative. It is conceivable 
that in future the two methods could be combined in order to increase the efficiency or for finding 
alternate routes for QIP.

\section{The Conditional phase shift gate}
  The conditional phase shift gate introduces a phase shift only if a 
predetermined condition is satisfied. In one qubit system, the conditional 
 phase shift gate is an unitary transform of the form  \cite{ic};\\
\begin{eqnarray}
 C_1(\phi)=\pmatrix{1&0 \cr
                              0&e^{i\phi}}  
\hspace{.3cm} {\mathrm or}
\hspace{.3cm}
 C_0(\phi)=\pmatrix{e^{i\phi}&0 \cr
                              0&1}
\end{eqnarray} 
 The $C_1(\phi)$ introduces a phase shift if the qubit is in state $\vert 1\rangle$ and 
 $C_0(\phi)$ introduces a phase shift if the qubit is in state $\vert 0\rangle$. However,
 these two operations are identical within an overall phase, since  
 $C_0(\phi)= e^{i\phi}C_1(-\phi)$.
 The phase shift gate can be realized experimentally  by a rotation of the 
 magnetization vector of a spin-1/2 nucleus (qubit) by angle $\phi$ about 
 z-axis (z-rotation). A $\phi$ angle pulse about z axis  has the form
\begin{eqnarray}
(\phi)_{z}=exp(-i\phi\sigma_z/2)=\pmatrix{e^{-i\phi/2}&0 \cr
                                 0&e^{i\phi/2}}
\end{eqnarray}
 where $\sigma_z$ is Pauli's z-matrix. We note that the phase gate $C_1(\phi)$  
 can be achieved using $(\phi)_{z}$, since;
\begin{eqnarray}
C_1(\phi)=e^{i\phi/2}(\phi)_z.
\end{eqnarray}
A $\phi$ angle pulse about z axis can be experimentally realized by a composite
 pulse of the form $(\phi)_z= (\pi/2)_y (\phi)_x (\pi/2)_{-y}$, where 
$(\phi)_y$ means a rotation of magnetization vector by an angle 
 $(\phi)$ about y-axis. This pulse sequence in widely known in NMR as a composite z-pulse\cite{lev,er}.
 
 In a two-qubit system there are 
four possible conditional phase shift 
gates ($C_{00}(\phi),C_{01}(\phi),C_{10}(\phi)$ and $C_{11}(\phi)$). A $C_{11}(\phi)$ 
gate introduces a phase shift of $\phi$ only if both the qubits are $\vert 1\rangle$, whereas a
 $C_{10}(\phi)$ gate does the same only if first qubit is $\vert 1\rangle$ and second qubit is $\vert 0\rangle$. 
 Similar logic holds for the other two gates. The unitary operator corresponding 
to $C_{11}(\phi)$ is 
\begin{eqnarray}
C_{11}(\phi)=\pmatrix{1&0&0&0 \cr
                 0&1&0&0 \cr
                 0&0&1&0 \cr
                 0&0&0&e^{i\phi}}.
\end{eqnarray}
 This gate has been implemented earlier using J-coupling evolution and 
spin selective pulses\cite{qft}. Here we demonstrate that $C_{11}(\phi)$ can also be 
constructed using transition selective z-pulses. For this purpose we first describe 
 phase rotation of a qubit.
The unitary operator describing a rotation of angle $\phi$ about z axis on the first qubit, 
when the second qubit is in state $\vert 0\rangle$ has the form
\begin{eqnarray}
(\phi)_{z0}&=&exp(-i\phi(\frac{1}{2}\sigma^{(1)}_z\otimes\sigma^{(2)}_0))\nonumber \\
           &=&\pmatrix{e^{-i\phi/2}&0&0&0 \cr
                 0&1&0&0 \cr
                 0&0&e^{i\phi/2}&0 \cr
                 0&0&0&1}.
\end{eqnarray}
 This is a transition-selective phase rotation by angle $\phi$ about z-axis.
 $\sigma^{(2)}_0$ is the polarization operator of the second qubit corresponding  
to the state $\vert 0\rangle$. The polarization operators of $j^{th}$ qubit when 
it is in state $\vert 0\rangle$ or $\vert 1\rangle$ are 
\begin{eqnarray}
\sigma^{(j)}_0 &=& \vert 0\rangle\langle0\vert = \pmatrix{1&0 \cr
                     0&0}\nonumber \\
\sigma^{(j)}_1 &=& \vert 1\rangle\langle1\vert =\pmatrix{0&0 \cr
                     0&1},
\end{eqnarray} 
 respectively\cite{er}. Similarly a $\phi$ angle pulse about z-axis on the first qubit when the second qubit 
is in state $\vert 1\rangle$ has the matrix form
\begin{eqnarray}
(\phi)_{z1}&=&exp(-i\phi(\frac{1}{2}\sigma^{(1)}_z\otimes\sigma^{(2)}_1)) \nonumber \\
           &=&\pmatrix{1&0&0&0 \cr
                 0&e^{-i\phi/2}&0&0 \cr
                 0&0&1&0 \cr
                 0&0&0&e^{i\phi/2}}. 
\end{eqnarray}
The $\phi$ angle z-rotation of second qubit when the first qubit is respectively in the state $\vert 0\rangle$ 
and $\vert 1\rangle$ are 
\begin{eqnarray}
(\phi)_{0z}&=&exp(-i\phi(\sigma^{(1)}_0\otimes\frac{1}{2}\sigma^{(2)}_z)) \nonumber \\
           &=&\pmatrix{e^{-i\phi/2}&0&0&0 \cr
                 0&e^{i\phi/2}&0&0 \cr
                 0&0&1&0 \cr
                 0&0&0&1},
\end{eqnarray}

\begin{eqnarray}
(\phi)_{1z}&=&exp(-i\phi(\sigma^{(1)}_1\otimes\frac{1}{2}\sigma^{(2)}_z)) \nonumber \\
          &=&\pmatrix{1&0&0&0 \cr
                 0&1&0&0 \cr
                 0&0&e^{-i\phi/2}&0 \cr
                 0&0&0&e^{i\phi/2}}. 
\end{eqnarray} 
The conditional phase shift gate $C_{11}(\phi)$ can now be realized by the sequence of 
z-rotations, $[(\phi/2)_{z0} (\phi/2)_{z1}][(\phi)_{1z}]$ with an 
overall phase of $e^{-i\phi/4}$, as
\begin{eqnarray} 
C_{11}(\phi)&=&[(\phi/2)_{z0} (\phi/2)_{z1}] [(\phi)_{1z}]\nonumber \\
             &=& e^{-i\phi/4}\pmatrix{1&0&0&0 \cr
                 0&1&0&0 \cr
                 0&0&1&0 \cr
                 0&0&0&e^{i\phi}}.          
\end{eqnarray}
  The $C_{11}(\phi)$ gate is also termed as Controlled phase shift gate\cite{ic}. 
Similarly, the other conditional phase gates in the two qubit system can be achieved within an 
overall phase of $e^{-i\phi/4}$ by the pulse 
sequences 
 \begin{eqnarray}
 C_{10}(\phi)&=&[(\phi/2)_{z0} (\phi/2)_{z1}] [(\phi)_{0(-z)}],\\
 C_{01}(\phi)&=&[(\phi/2)_{(-z)0}(\phi/2)_{(-z)1}][(\phi)_{1z}],\\
 C_{00}(\phi)&=&[(\phi/2)_{(-z)0}(\phi/2)_{(-z)1}][(\phi)_{0(-z)}].
\end{eqnarray}
 The pulses in the first bracket are transition selective pulses on the
transitions of first qubit, while the second bracket has a transition selective pulse
on a transition of second qubit.
Each pulse about z-axis is experimentally realized by transition selective composite z-pulses.
 In a three qubit system, there are eight conditional phase shift
 gates($C_{000}(\phi),C_{001}(\phi),...,C_{111}(\phi)$). 
The $C_{111}(\phi)$ gate introduces a phase shift of $\phi$ when all the three qubits are in state $\vert 1\rangle$
 and does nothing otherwise.
 The unitary operator of the $C_{111}(\phi)$ gate is 
\begin{equation}
\hspace*{1cm}  C_{111}(\phi)=\pmatrix{1&0&0&0&0&0&0&0 \cr
                                  0&1&0&0&0&0&0&0 \cr
                                  0&0&1&0&0&0&0&0 \cr
                                  0&0&0&1&0&0&0&0 \cr
                                  0&0&0&0&1&0&0&0 \cr
                                  0&0&0&0&0&1&0&0 \cr
                                  0&0&0&0&0&0&1&0 \cr
                                  0&0&0&0&0&0&0&e^{i\phi}}.
\end{equation} 
With the same logic as applied to one and two qubit systems, we realize the phase shift gate 
$C_{111}(\phi)$(with an overall phase factor of $e^{-i\phi/8}$) by a sequence of transition selective z-pulses,
\begin{eqnarray}
 C_{111}(\phi)&=& [(\phi/4)_{z00}  (\phi/4)_{z01}  (\phi/4)_{z10} 
               (\phi/4)_{z11}]\nonumber \\ &&~~~~[(\phi/2)_{1z0 } (\phi/2)_{1z1}] [ (\phi)_{11z}].
\end{eqnarray}
The pulses in the first bracket are on first qubit, second bracket on second qubit and 
third bracket on third qubit.
The other phase shift gates can be achieved by the same number of transition selective 
z-rotations, with different pulses on transitions of second and third qubits. 
For example two other phase gates $C_{000}(\phi)$ 
and $C_{110}(\phi)$, which will be used later for demonstration of Grover's search algorithm, are
\begin{eqnarray}
 C_{110}(\phi)&=& [(\phi/4)_{z00}  (\phi/4)_{z01}  (\phi/4)_{z10} \nonumber
(\phi/4)_{z11}] \\&&~~~~[(\phi/2)_{1z0 } (\phi/2)_{1z1}] [(\phi)_{11(-z)}],\\
 C_{000}(\phi)&=& [(\phi/4)_{(-z)00}  (\phi/4)_{(-z)01}  (\phi/4)_{(-z)10} \nonumber 
 (\phi/4)_{(-z)11}]\\&&~~~~[(\phi/2)_{0(-z)0 } (\phi/2)_{0(-z)1}] [(\phi)_{00(-z)}].
\end{eqnarray}
Like $C_{111}(\phi)$ case, these pulse sequences also have an overall phase of $e^{-i\phi/8}$. 
 
 The above sequences can be easily generalized into a sinlge formula to build an N-qubit 
 conditional phase shift gate $C_{ijk...mn}(\phi)$ (where the N qubits are  
 $i,j,k,...m,n$= 0 or 1), as given below; 
\begin{eqnarray}
&&C_{ijk...mn}(\phi)=\nonumber\\ 
&&\left[\prod_{j'k'...m'n'}^{2^{N-1}}(\frac{\phi}{2^{N-1}})_{\{(-1)^{i+1}z\}j'k'...m'n'}\right] 
\nonumber \\
                    &&\left[\prod_{k'...m'n'}^{2^{N-2}}(\frac{\phi}{2^{N-2}})_{i\{(-1)^{j+1}z\}k'...m'n'}\right]... 
\nonumber \\
                    &&... \left[\prod_{n'}^{2}(\frac{\phi}{2})_{ijk...\{(-1)^{m+1}z\}n'}\right] 
                       \left[(\phi)_{ijk..m\{(-1)^{n+1}\}z}\right],	
\end{eqnarray}
 where $j',k',...m',n'$=0 or 1. It is to noted that the above Eqs.~[10]-[18] are not unique. 
The smallest angle pulse (${\phi}/{2^{(N-1)}}$) need not be applied only on the first qubit but 
can be applied on any qubit. Similarly pulse (${\phi}/{2^{(N-2)}}$) can be applied on any qubit 
other than on the qubit on which (${\phi}/{2^{(N-1)}}$) is applied. Hence different combinations 
of transition selective z-pulses can create the same gate, but in all pulse sequences the logic of
 Eq.~[18] is maintained. For example, the $C_{111}(\phi)$ given in Eq.~[15] can also be created by  
 different pulse sequences such as;
\begin{eqnarray}
 C_{111}(\phi)&=& [(\phi/4)_{0z0}  (\phi/4)_{1z0}  (\phi/4)_{0z1}
               (\phi/4)_{1z1}]\nonumber \\ &&~~~~[(\phi/2)_{z10 } (\phi/2)_{z11}] [ (\phi)_{11z}]\nonumber \\
              &=& [(\phi/4)_{00z}  (\phi/4)_{10z}  (\phi/4)_{01z}
               (\phi/4)_{11z}]\nonumber \\ &&~~~~[(\phi/2)_{0z1} (\phi/2)_{1z1}] [ (\phi)_{z11}]\nonumber \\
              &=& [(\phi/4)_{00z} (\phi/4)_{10z}  (\phi/4)_{01z}
               (\phi/4)_{11z}]\nonumber \\ &&~~~~[(\phi/2)_{z01} (\phi/2)_{z11}] [ (\phi)_{1z1}].
\end{eqnarray}
 However, all the sequences require the same number of pulses.
 
 For quantum Fourier transform in a three qubit system, a reduced conditional phase shift gate 
is required where the condition is on two qubits and there is no condition on the third qubit. 
For example, $C_{11\epsilon}(\phi)$ gate acts according to the states of 
first and second qubits and introduces a phase shift only if both the qubits 
are in state $\vert 1\rangle$ (shown by subscript), and is independent of the state of the  
 third qubit which can be in state $\vert 0\rangle$ or $\vert 1\rangle$ ($\epsilon$=0 or 1). 
The unitary operator of $C_{11\epsilon}(\phi)$ gate is
\begin{equation}
  C_{11\epsilon}(\phi)=\pmatrix{1&0&0&0&0&0&0&0 \cr
                                  0&1&0&0&0&0&0&0 \cr
                                  0&0&1&0&0&0&0&0 \cr
                                  0&0&0&1&0&0&0&0 \cr
                                  0&0&0&0&1&0&0&0 \cr
                                  0&0&0&0&0&1&0&0 \cr
                                  0&0&0&0&0&0&e^{i\phi}&0 \cr
                                  0&0&0&0&0&0&0&e^{i\phi}}.
\end{equation}
 The pulse sequences of the reduced conditional phase shift gate can be easily constucted from 
the pulse sequences of the conditional phase shift gates. 
For example, 
\begin{equation}
  C_{11\epsilon}(\phi)=C_{111}(\phi)C_{110}(\phi)
\end{equation}
 Using Eqs.~[15],[16] and [21], 
 this gate can be realized (with an overall phase of $e^{-i\phi/4}$)
 by a sequence of transition selective z-pulses;
\begin{eqnarray}
 C_{11\epsilon}(\phi)&=&[(\phi/4)_{z00}  (\phi/4)_{z01}  (\phi/4)_{z10}
               (\phi/4)_{z11}]\nonumber \\ &&~~~~[(\phi/2)_{1z0 } (\phi/2)_{1z1}] [ (\phi)_{11z}]\times \nonumber \\
                     && [(\phi/4)_{z00}  (\phi/4)_{z01}  (\phi/4)_{z10}
               (\phi/4)_{z11}]\nonumber \\ &&~~~~[(\phi/2)_{1z0 } (\phi/2)_{1z1}] [ (\phi)_{11(-z)}].
\end{eqnarray}
 Since all z-pulses commute Eq.~[22] reduces to;
\begin{eqnarray}
 C_{11\epsilon}(\phi)&=& [(\phi/4)_{z00}  (\phi/4)_{z01}  (\phi/4)_{z10}
               (\phi/4)_{z11}]^2 \nonumber \\ &&[(\phi/2)_{1z0 } (\phi/2)_{1z1}]^2[(\phi)_{11z}(\phi)_{11(-z)}]. 
\end{eqnarray}
 The last two pulses cancel and the squares of exponential operators double the angle of rotation, yielding,
\begin{eqnarray}
 C_{11\epsilon}(\phi)&=& [(\phi/2)_{z00}  (\phi/2)_{z01}  (\phi/2)_{z10}
                          (\phi/2)_{z11}]\nonumber \\&&~~~~ [(\phi)_{1z0 } (\phi)_{1z1}]. 
\end{eqnarray}
This reduced gate requires one z-pulse less than $C_{111}(\phi)$ or $C_{110}(\phi)$ gate.
 Other reduced conditional phase shift gates can also be similarly realized (with an overall phase of $e^{-i\phi/4}$) as:
\begin{eqnarray}
 C_{1\epsilon1}(\phi)&=& [(\phi/2)_{z00}  (\phi/2)_{z01}  (\phi/2)_{z10}
(\phi/2)_{z11}] \nonumber \\&&~~~~[(\phi)_{10z } (\phi)_{11z}],\\
  C_{\epsilon11}(\phi)&=& [(\phi/2)_{z00}  (\phi/2)_{z01}  (\phi/2)_{z10}
(\phi/2)_{z11}] \nonumber \\&&~~~~ [(\phi)_{01z } (\phi)_{11z}].
\end{eqnarray}  
 Thus, these gates can also be realized by transition selective pulses. 
One can extend the composite transition selective z-pulse sequence to construct a
`m'-qubit conditional phase gate in any `N' qubit system (with an overall phase of $e^{(-i\phi/2^m)}$). 
For such a gate, the number of 
transition selective z-pulse required are (2$^N$-2$^{N-m}$). The number of pulses 
 decreases as the number of conditional qubits `m' decrease. Each transition selective z-pulse 
 is experimentally realized by three transition selective r.f. pulses along x and y axis\cite{ran}. As 
an example the z-pulse on one of the transitions of third qubit in a N-qubit system can be realized as
\begin{eqnarray}
 (\phi)_{ijzl...n}&=&(\pi/2)_{ijyl...n} (\phi)_{ijxl...n}  \nonumber
\\&& (\pi/2)_{ij-yl...n}
\end{eqnarray} 
where $i,j,l,...n$=0 or 1. 

 It may be mentioned that for all the gates described above (Eqs.[10]-[27]) the first set of 
$2^{N-1}$ transition selective z-pulses act on all transitions of a spin (qubit), and 
wherever possible (in weakly coupled spin-1/2 systems), can be experimentally implemented 
by using a spin (qubit) selective z-pulse, thus reducing the number of transition selective pulses by 
2$^{N-1}$ pulses. For example, the C$_{11\epsilon}$ type gates (Eqs.[24]-[26]) require one spin selective 
and two transition selective pulses.   
\section{Simulations}
\subsection{Grover's search algorithm}
 Grover's search algorithm can search (with high probability) any state of an N-qubit system in
 $O(\sqrt{2^N})$ iterations\cite{gr}. 
 Each iteration has two steps, namely `conditional sign-flip' and `inversion about average'. 
 These can be implemented by using the conditional phase shift gate. We demonstrate here the implementation of Grover's 
search algorithm by simulation using transition selective pulses on a 3 qubit system.
 In a 3-qubit system, the algorithm requires 2 iterations. The algorithm starts with 
an initial pseudo-pure state, say $\vert000\rangle$. A Hadamard gate `H' is applied on this state. 
The Hadamard gate 
rotates each qubit from $\vert 0\rangle$ state to an uniform superposition $(\vert 0\rangle+\vert 1\rangle)/\sqrt{2}$. 
The unitary transform of Hadamard gate for $j^{th}$ qubit is \cite{pre}
\begin{eqnarray}
 H_{j}=\frac{1}{\sqrt{2}}\pmatrix{1&1 \cr
                                   1&-1},
\end{eqnarray}
 and when applied on all three qubits, \b it is of the form 
\begin{eqnarray}
    H&=& H_1\otimes H_2\otimes H_3 \nonumber \\&=& \frac{1}{\sqrt{2}}\pmatrix{1&1 \cr
                                   1&-1} \otimes \frac{1}{\sqrt{2}}\pmatrix{1&1 \cr
                                   1&-1} \otimes \frac{1}{\sqrt{2}}\pmatrix{1&1 \cr
                                   1&-1}.
\end{eqnarray}
 The Hadamard gate on all qubits, applied on a pseudo-pure state, 
creates an uniform superposition of all possible states,
\begin{eqnarray}
&& \vert \psi_1\rangle= H \vert 000\rangle=\nonumber
\\&& \frac{1}{2\sqrt{2}}\pmatrix{1&1&1&1&1&1&1&1 \cr
                      1&-1&1&-1&1&-1&1&-1 \cr
                      1&1&-1&-1&1&1&-1&-1 \cr
                      1&-1&-1&1&1&-1&-1&1 \cr
                      1&1&1&1&-1&-1&-1&-1 \cr
                      1&-1&1&-1&-1&1&-1&1 \cr
                      1&1&-1&-1&-1&-1&1&1 \cr
                      1&-1&-1&1&-1&1&1&-1}
                          \pmatrix{1 \cr 0 \cr 0 \cr 0 \cr
                                     0 \cr 0 \cr 0 \cr 0 \cr} \nonumber \\
                        &=&\frac{1}{2\sqrt{2}} \pmatrix{1 \cr 1 \cr 1 \cr 1 \cr
                                    1 \cr 1 \cr 1 \cr 1 \cr}.
\end{eqnarray}
 $H$ gate is realized experimentally by the pulse sequence $(\pi/2)_{-y} (\pi)_{x}$ 
(hard pulses, applied on all qubits).
  Grover's iteration starts from this point by the application of `conditional sign-flip'.
In the `conditional sign-flip' step, 
the state which is being searched is inverted; that is  a phase shift of $\phi= \pi$ is introduced 
in that particular state. This can be achieved by a conditional phase shift gate with $\phi= \pi$ 
corresponding to that state. 
For example, for the search of $\vert 110\rangle$ state, the conditional sign flip is the conditional phase 
shift gate $C_{110}(\pi)$. The same logic holds for searches of other states.   

 The second step of Grover's iteration is `inversion about average', 
 in which all the states are inverted  about their average amplitude. 
 The unitary operator ($\Lambda$) of this step for a 3-qubit system 
 is of the form \cite{gr} 
\begin{equation}
\Lambda= \frac{1}{4}\pmatrix{-3&1&1&1&1&1&1&1 \cr
                         1&-3&1&1&1&1&1&1 \cr
1&1&-3&1&1&1&1&1 \cr
1&1&1&-3&1&1&1&1 \cr
1&1&1&1&-3&1&1&1 \cr
1&1&1&1&1&-3&1&1 \cr
1&1&1&1&1&1&-3&1 \cr
1&1&1&1&1&1&1&-3}.
\end{equation}
This step can be realized by applying a Hadamard gate `$H$' (on all qubits) before and 
after a conditional phase shift $C_{000}(\pi)$ gate, hence $\Lambda=HC_{000}(\pi)H$\cite{gr,ci}. 
 It may be noted that the $C_{000}(\pi)$ gate is required because we started with 
  $\vert 000\rangle$ pseudo-pure state, whereas if one starts with another pseudo-pure state, say
 $\vert 001\rangle$, then the corresponding phase shift gate $C_{001}(\pi)$ will be required.  
After one full Grover's iteration (starting from $\vert 000\rangle$ pseudo-pure state and searching for 
  $\vert 110\rangle$ state), the state of the system is 
\begin{eqnarray}
 \vert \psi_2\rangle&=&(\Lambda) (C_{110}(\pi)) \vert \psi_1\rangle\nonumber \\
             \vspace{-8cm}
             &=&(HC_{000}(\pi)H)( C_{110}(\pi))\vert \psi_1\rangle
                         =\frac{1}{4\sqrt{2}} \pmatrix{1 \cr 1 \cr 1 \cr 1 \cr
                                    1 \cr 1 \cr 5 \cr 1 \cr}.
\end{eqnarray} 
 After the second Grover's iteration (conditional sign-flip and inversion about average) we 
find the final state to be 
\begin{eqnarray}
 \vert \psi_3\rangle&=&[\Lambda C_{110}(\pi)] \vert \psi_2\rangle \nonumber \\
             &=&[\Lambda C_{110}(\pi)][\Lambda C_{110}(\pi)]\vert \psi_1\rangle \nonumber
             \\&=&[HC_{000}(\pi)H C_{110}(\pi)]^2 H\vert 000\rangle \nonumber
              \\           &=&\frac{1}{8\sqrt{2}} \pmatrix{-1 \cr -1 \cr -1 \cr -1 \cr
                                    -1 \cr -1 \cr 11 \cr -1 \cr}.
\end{eqnarray}
The result of measurement on the final state will give the searched state $\vert 110\rangle$ 
with high probability. The entire sequence for two iterations after pseudo pure state is
$HC_{000}(\pi)HC_{110}(\pi)HC_{000}(\pi)HC_{110}(\pi)H$. The pulse sequence for 
the required phase shift gates $C_{110}$ and $C_{000}$ are as in Eqs.[16] and [17], which use 
transition selective pulses. This demonstrates that the Grover's search algorithm can be 
implemented using transition selective pulses
complementing the coupling evolution method proposed earlier\cite{ap}. 
Fig.~1 shows the density matrices of the 
system at (a) initial pseudo-pure state, followed by (b) state of uniform superposition (Eq. [30]),
 (c) after conditional sign-flip, (d) inversion about average (Eq. [32]), (e) after conditional sign-flip of 
second iteration, and (f) after inversion about average of second iteration which is the final result (Eq. [33]). 
 The searched state is clearly identified in Fig.~1(f).
\subsection{Quantum Fourier transform}
  Just as classical Fourier transform extracts periodicity in functions, 
quantum Fourier transform (QFT) extracts periodicity of wave functions. It is defined as follows:
\begin{equation}
  QFT_q\vert x\rangle=\frac{1}{\sqrt{q}} \sum_{x'=0}^{q-1} e^{2\pi i xx'/q} \vert x'\rangle
\end{equation}
where $q=2^n$ is the dimension of Hilbert space for a n-qubit system. If $f(x)$ is periodic 
with periodicity r, then the corresponding quantum Fourier transformed function $f(p)$ will give 
a peak at $p=q/r$. The quantum circuit for three qubit QFT \cite{ic} is given in Fig.~2. 

  Quantum Fourier transform has been demonstrated using J-coupling evolution by 
Weinstein et.al.\cite{qft}. Here we demonstrate quantum Fourier transform using transition selective 
pulses. From quantum circuit, Fig.~2, one can infer the QFT gate sequence in three qubit system as
\begin{eqnarray}
QFT_8= SWAP_{13}H_3C_{\epsilon11}(\pi/2)C_{1\epsilon1}(\pi/4)H_2C_{11\epsilon}(\pi/2)H_1. 
\end{eqnarray}
$H_j$ is the Hadamard gate 
operated on $j^{th}$ qubit. Here the operation are applied in the sequence from right to left 
such that the $SWAP_{13}$ is the last operation. The pulse sequence of $C_{11\epsilon},C_{1\epsilon1}$ and 
$C_{\epsilon11}$ 
gates are given in Eqs.~[24]-[26]. The $SWAP_{13}$ gate performs a swap between qubits 1 and 3, 
and is realized by a cascade of transition selective $\pi$ pulses \cite{lo}. 
\begin{equation}
 SWAP_{13}:[(\pi)_{00x} (\pi)_{x00 } (\pi)_{00x}]
             [(\pi)_{11 x} (\pi)_{x11} (\pi)_{11x}].
\end{equation}
 It may be noted that in Eq. [36] the first set of three pulses taken together commute with the last 
set of three pulses taken together, and can be applied in any order.   
 QFT sequence will extract the periodicity of an input state of any periodicity r. We 
demonstrate QFT for two different inputs. First input has state periodicity $r=4$, and is created by a 
$(\pi/2)_y$ qubit selective hard pulse on the first qubit, Fig. 3(a). 
The output state obtained by applying sequence of Eq. [35] shows a periodicity of $q/r=2^3/4=2$, 
as is evident from the density matrix of 
output state, shown in Fig. 3(b). The second input state has periodicity 2, and is created by a
$(\pi/2)_y$ qubit selective hard pulse on the first and second qubit, Fig.~3(c). The output state 
obtained by applying sequence of Eq. [35], has expected 
periodicity of $q/r=4$ as shown in Fig.3(d). 

\section{Experiments}
 Grover's 2-qubit search has also been carried out experimentally using transition selective pulses on a
two qubit system formed by the carbon-13 and protons of the molecule 
 $^{13}$CHCl$_{3}$. Experiment have been performed at room temperature in 
DRX 500 spectrometer. The coherence times were; $T_1$=20 sec and $T_2$=0.4 sec for the 
proton, $T_1$=21 sec and $T_2$=0.3 sec for the carbon. The proton resonance frequency on the DRX 500 MHz 
spectrometer is 500.13 MHz, and that of carbon-13 is 125.76 MHz. The spin-spin (J) coupling 
in this system is 209 Hz.
  Gaussian shaped pulses of 
duration 20ms were used as transition selective pulses. 
 The initial pseudo-pure state were also prepared using transition selective pulses. A pulse sequence 
[$(70.5 ^o)_{x 1} (90^o)_{1 x}-grad$] equalizes the populations of the states $\vert 01\rangle$,
$\vert 10\rangle$ and $\vert 11\rangle$,
keeping the population of $\vert 00\rangle$ state undisturbed, different from the equalized population of other states; 
 hence establishing $\vert 00\rangle$ pseudo-pure state\cite{ernst,mah}. An inhomogeneous magnetic field gradient 
pulse along z-direction destroys any created coherence. 
This state is shown in Fig. 4(a).

 Grover's algorithm for 2-qubit system requires only one iteration.
The gate sequence (after pseudo-pure state) for two qubit search is 
$HC_{ij}(\pi)HC_{00}(\pi)H$, where ij= 00, 01, 10, and 11 for searching the states $\vert 00\rangle$, 
$\vert 01\rangle$, $\vert 10\rangle$, and $\vert 11\rangle$ 
respectively. The first Hadamard $H$ gate is applied to create uniform superposition state, 
 the  $C_{ij}$ phase gate is for `conditional sign-flip step', and $HC_{00}(\pi)H$ performs 
 the `inversion about average'.
The pulse sequence used for different $C_{ij}$ phase gates are as in Eqs. [10]-[13]. 
 Each $C_{ij}$ requires three z-pulses. The first two z-pulses are to be applied on both the transitions 
of first qubit, hence making it a qubit selective z-rotation.
\begin{eqnarray}
 (\phi/2)_{z0} (\phi/2)_{z1}= (\phi/2)_{z}^{(1)},
\end{eqnarray}
 where the superscript shows that it is a z-rotation of qubit 1.
 The large Larmour frequency difference between two qubits ($^{13}$C and $^1$H) 
allows one to achieve the first two z-pulses by qubit selective hard pulses on the first qubit.
 For example, in the case of $C_{11}(\phi)$ gate 
\begin{eqnarray}
 [(\phi/2)_{z0}] [(\phi/2)_{z1}] &=& [(\pi/2)_{y0} (\phi)_{x0}(\pi/2)_{(-y)0}] \nonumber \\ 
                                   && [(\pi/2)_{y1} (\phi)_{x1}(\pi/2)_{(-y)1}] \nonumber \\
                                 &=&(\pi/2)_{y}^{(1)} (\phi)_{x}^{(1)} (\pi/2)_{(-y)}^{(1)}  
\end{eqnarray}
  The third z-pulse $\phi_{1z}$ was achieved 
by three transition selective pulses on a transition of the second qubit, as in case of $C_{11}(\phi)$, 
$(\phi)_{1z}= (\pi/2)_{1y} (\phi)_{1x}(\pi/2)_{1(-y)}$. Hence each 
conditional phase shift gate required three qubit selective and three transition selective pulses.    
  We have experimentally determined the 
search of all the states $\vert 00\rangle$,$\vert 01\rangle$,$\vert 10\rangle$ and 
$\vert 11\rangle$ in four different experiments,
 the results being shown in Fig. 4. 

 The pseudo-pure state and the final states were tomographed by 
efficient quantum state tomography using two-dimensional Fourier transform technique\cite{ran}.
The diagonal elements were measured using a one-dimensional experiment. 
In this experiment all 
off-diagonal elements were dephased by a gradient pulse
 along the z-direction, and then a 
small angle $(15^o)$ pulse was used for detection. The 
detected signal yielded all the diagonal elements. The off-diagonal 
elements were measured using a two-dimensional experiment. Here the density matrix was 
allowed to evolve for a time  $t_1$, then a $(90^o)$ hard pulse (on all qubits) 
transforms every element of 
density matrix into all other elements including diagonal elements. A gradient pulse retains 
only diagonal elements, and a  $(45^o)$ hard pulse (on all qubits) 
is applied and the signal is detected as a function of time $t_2$.
 The detected signal is Fourier transformed with respect to $t_1$ and $t_2$, yielding 
a two-dimensional frequency domain spectrum S($\omega_1,\omega_2$). 
 The intensities of various peaks in S($\omega_1,\omega_2$) yields all the 
 off-diagonal elements of the initial density matrix. 
The above procedure of measuring off-diagonal elements requires 
ideal $(90^o)$ and $(45^o)$ pulse. Therefore we also performed 
another one-dimensional experiment without applying any pulses, which allows us to directly measure the single 
quantum elements of the initial density matrix. These intensities were compared with the single quantum elements 
measured by the two-dimensional experiment and were used to normalize the remaining off-diagonal 
elements.
 The errors in the diagonal elements were less than  $5\%$. 
 The errors in the off-diagonal elements were less than $15\%$. The errors
 are mainly due to errors in long low power rf pulses (transition selective), during 
 which relaxation and evolution adds to the errors. However, the errors in the present 
experiment are comparable to those performed earlier using J-evolution\cite{ci}.
\section{Conclusion}
 In this work we suggest transition selective pulses as candidate for implementation 
 of phase shift gates for quantum information 
processing. The implementation of unitary operators by J-evolution requires
refocusing pulse schemes for removal of evolution under chemical shifts and 
unwanted couplings\cite{jm}. The use
of transition selective pulses do not need such pulse schemes. 
 The search of more qubits has led researchers to use molecules
 oriented in liquid crystalline matrices as computers\cite{fu3,fu1,fu,ak}. In these systems, one often 
encounters spins which are strongly coupled, and in such cases evolution under J-coupling or
dipolar coupling  becomes too complex for quantum information processing.
\cite{abab}. The use of transition selective pulses for quantum information processing is
especially useful in such systems \cite{mah}.
 The use of 
low power, long duration r.f. pulses, gives rise to experimental errors due to relaxation 
and unwanted evolution under the internal Hamiltonian during long pulses. The experimental 
error may be reduced by using strongly modulating selective pulses of shorter duration,
 developed recently\cite{mod}.      
\section{Acknowledgments}
 Useful discussions with Prof. K.V. Ramanathan and Mr. Neeraj Sinha are 
 gratefully acknowledged. The use of DRX-500 NMR spectrometer of the Sophisticated 
Instuments Facility, Indian Institute of Science, Bangalore, is also gratefully acknowledged.

%*********************

* Author to whom correspondence should be addressed.
e-mail: $\it{anilnmr@physics.iisc.ernet.in}$
\end{multicols}

\pagebreak
\begin{figure}
\vspace{-1.5cm}
\epsfig{file=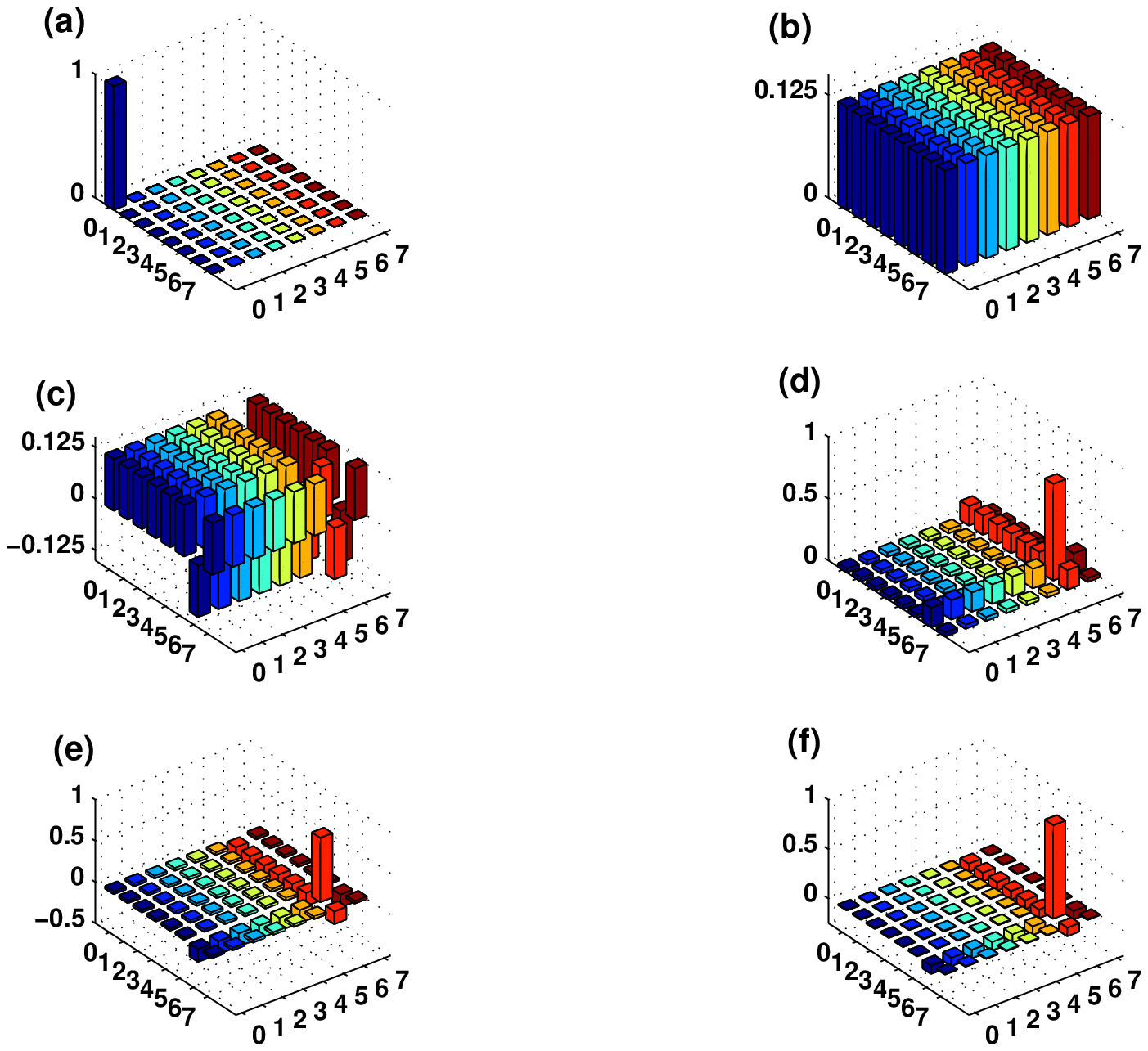,height=18cm,clip=,bbllx=100,bblly=200,bburx=300,bbury=600}
\end{figure}

{\small Fig.~1  Implementation of 3-qubit Grover's search algorithm is shown by simulations. The state $\vert 110\rangle$
is searched. The density matrices at different stages of the algorithm are shown. (a) Pseudopure $\vert 000\rangle$ state,
(b) Uniform superposition, (c) Conditional sign-flip, (d) Inversion about average, (e) Conditional sign-flip
of second iteration, and (f) Inversion about average of the second iteration. The high probablity of 
$\vert 110\rangle$ state
is reflected in the final density matrix. The x-axis and y-axis labels correspond to different states as:
 $0\rightarrow \vert 000\rangle,1\rightarrow \vert 001\rangle,2\rightarrow \vert 010\rangle,
3\rightarrow \vert 011\rangle,4\rightarrow \vert 100\rangle,
5\rightarrow \vert 101\rangle,6\rightarrow \vert 110\rangle$ and $7\rightarrow \vert 111\rangle$. }
\begin{figure}
\vspace{-22cm}
\hspace{7cm}
\epsfig{file=sim33.eps,height=18cm,clip=,bbllx=360,bblly=200,bburx=530,bbury=600}
\end{figure}

\pagebreak
\begin{center}
\begin{figure}
\epsfig{file=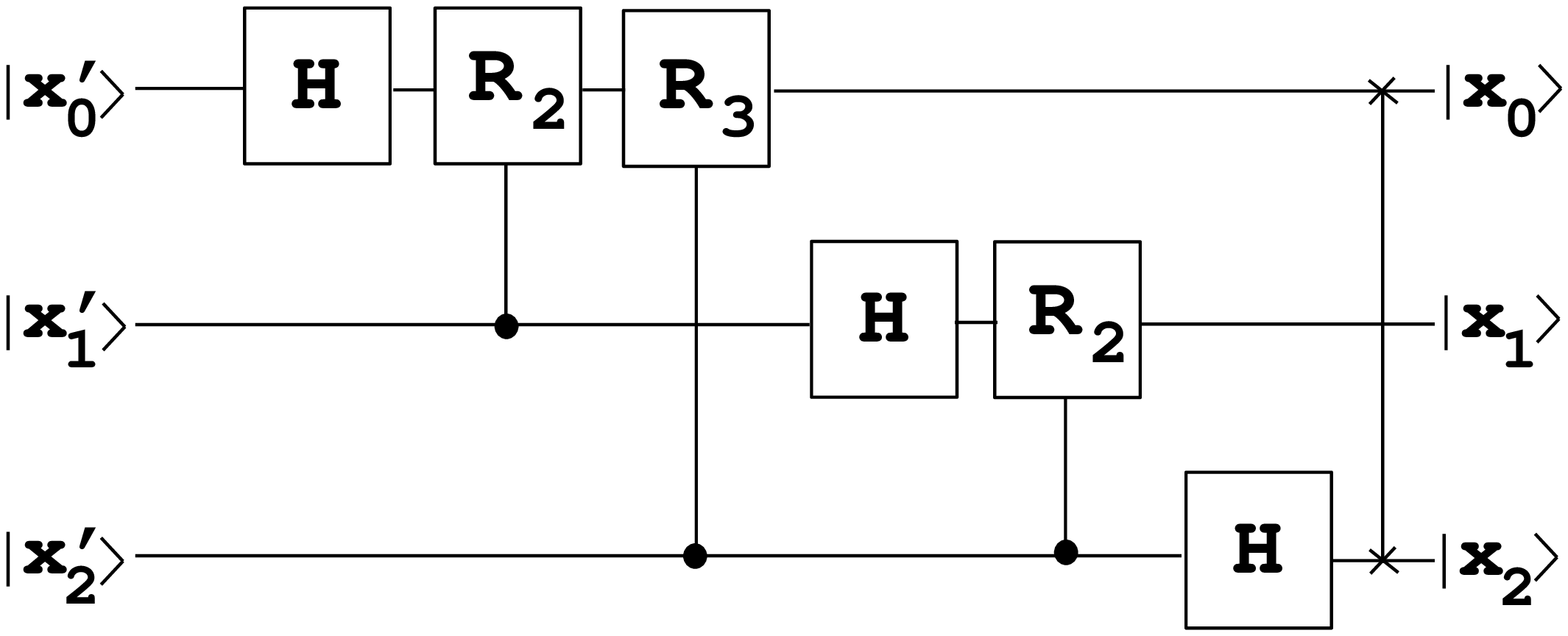,height=15cm,angle=270}
\end{figure}
\end{center}
\small {Fig.~2. Efficient quantum circuit for quantum Fourier transform in a three qubit system.
 $x_0',x_1'$ and $x_2'$ are states of the three qubits in the input, and, $x_0,x_1$ and $x_2$ are
the corresponding states in the output. The last operation is a swap gate between qubits 1 and 3.
 The unitary transformation $R_k$ is the phase gate $R_k=\pmatrix{1&0 \cr
             0&e^{2\pi i/2^k}}.$}

\pagebreak
\begin{figure}
\hspace{-2cm}
\epsfig{file=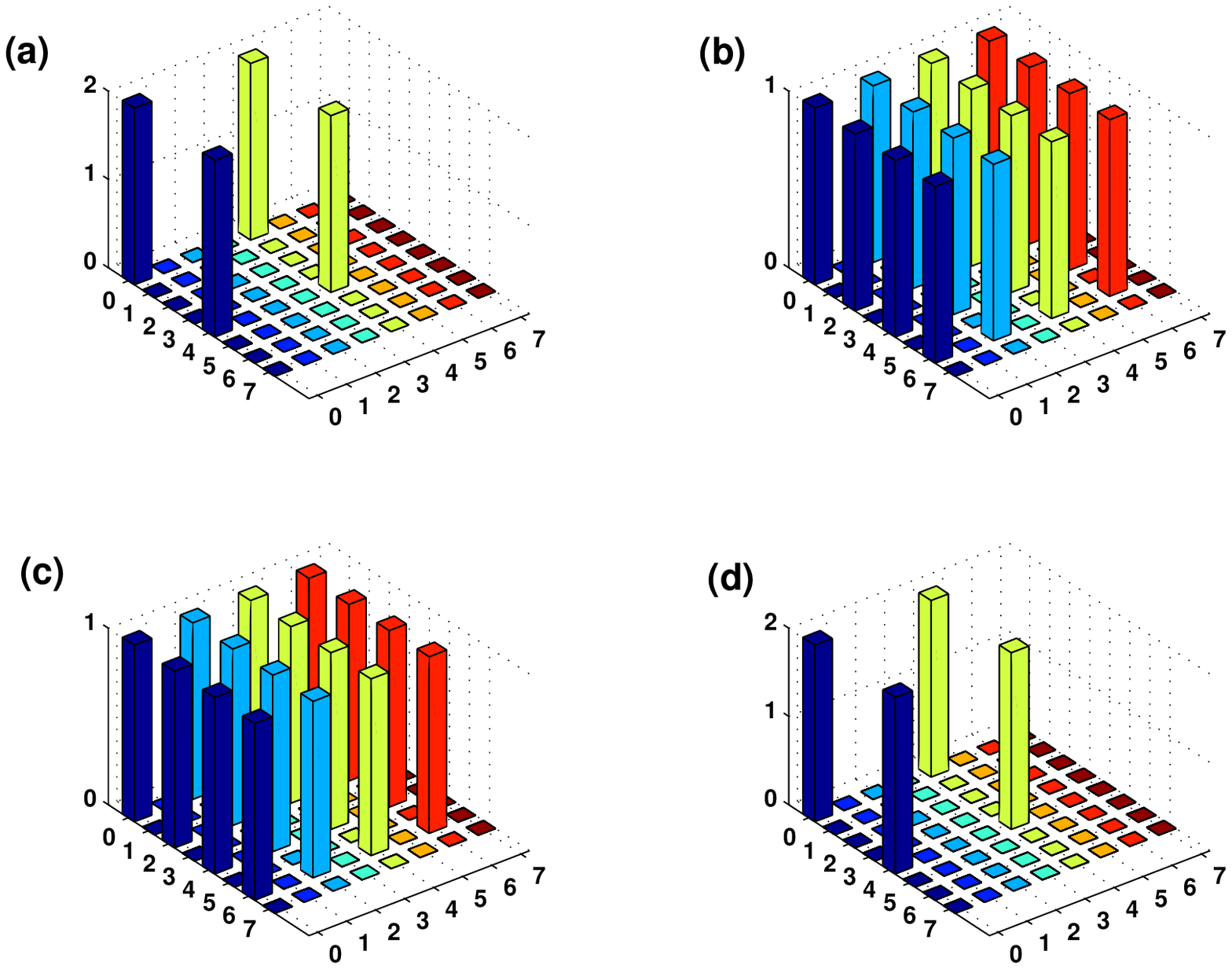}
\end{figure}
{\small Fig~3. Quantum Fourier transform in 3-qubits is shown by simulations. The dimension of the Hilbert's 
space of the system is 8. (a) and (c) are two different input states with periodicity (r) 4 and 2 respectively.
 The density matrix of the corresponding output states (b) and (d), show periodicities (q/r) as 2 and 4 respectively. 
 This shows QFT extracted the periodicity of the input state.The x-axis and y-axis labels correspond to different states as:
 $0\rightarrow \vert 000\rangle,1\rightarrow \vert 001\rangle,2\rightarrow \vert 010\rangle,
3\rightarrow \vert 011\rangle,4\rightarrow \vert 100\rangle,
5\rightarrow \vert 101\rangle,6\rightarrow \vert 110\rangle$ and $7\rightarrow \vert 111\rangle$.}
\pagebreak

\begin{figure}
\hspace{3.5cm}
\epsfig{file=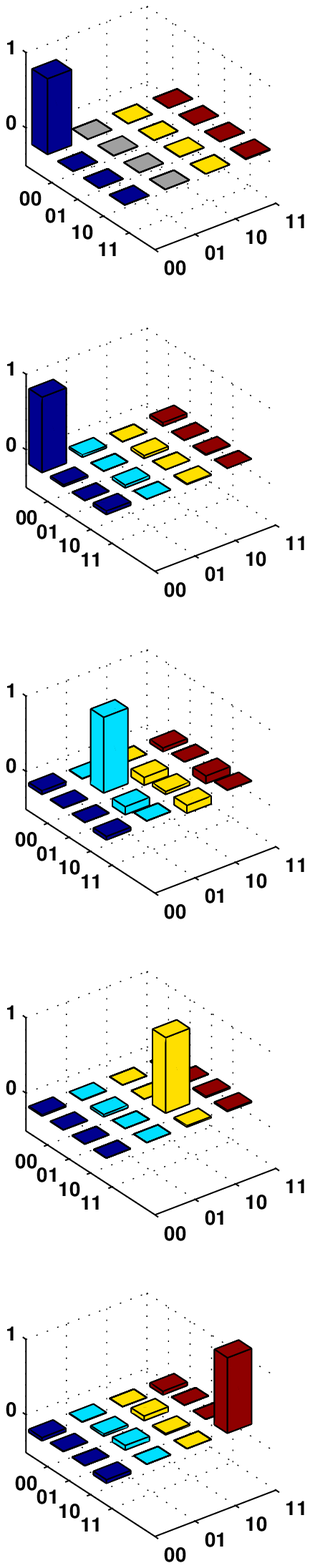,height=19.5cm,width=10cm}
\end{figure}
\small { Fig.~4. Experimental implementation of Grover's search algorithm on atwo qubit system using transition 
selective pulses.
(a) is the initial pseudopure state $\vert 00\rangle$, (b) is the search of $\vert 00\rangle$ state, 
(c) of $\vert 01\rangle$ state, (d) of $\vert 10\rangle$ state and
(e) is of $\vert 11\rangle$ state. The left hand figures show the spectra for mapping of the diagonal elements.
An inhomogenous magnetic field along z-direction kills all unwanted off-diagonal elements
created by pulse imperfections, and then a small angle $(15^o)$ pulse yielded the above spectra.
The real part of the complete tomographed density matrices are shown in the right hand side.}
\begin{figure}
\vspace{-22cm}
\hspace{2.5cm}
\epsfig{file=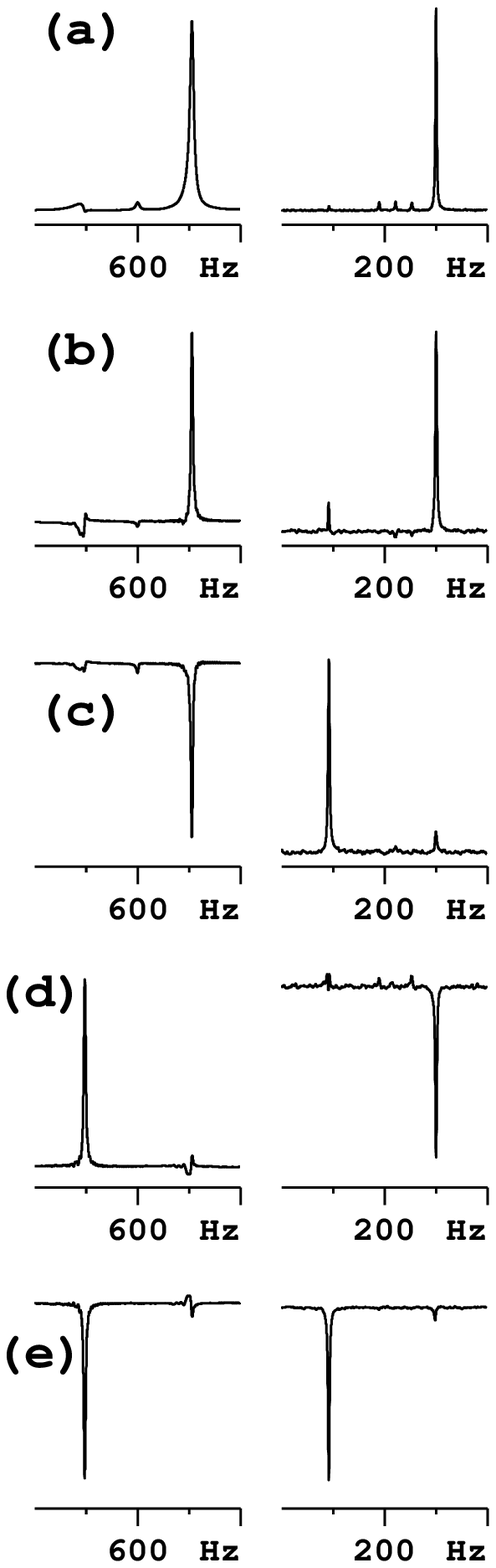,height=18cm}
\end{figure}


\begin{thebibliography}{99}
\bibitem{rf} R.P. Feynman, Simulating Physics with Computers. {\it Int. J. Theor. Phys.} {\bf 21} (1982) 467.
\bibitem{deu} D. Deutsch and R. Jozsa, Rapid solutions of problems by quantum computaion, 
{\it Proc. R. Soc. Lond. A} {\bf 454}, 339-354 (1998).
\bibitem{pw} P. W. Shor, Polynomial-time algorithms for prime factorization and discrete algorithms 
on quantum computer, {\it SIAM Rev}. {\bf 41}, 303-332 (1999).
\bibitem{ss} S. Lloyd, Universal Quantum Simulators, {\it Science.} {\bf 273} (1996) 1073.
\bibitem{gr} L.K. Grover, Quantum mechanics helps in searching for a needle in a haystack, 
{\it Phys. Rev. Lett.} {\bf 79}, 325 (1997).
\bibitem{pre} J.Preskill, "Lecture notes on quantum computation", http://
 theory.caltech.edu/people/preskill/ph229/
\bibitem{db} D. Bouwnmeester,A. Ekert,A. Zeilinger(Eds.), "The Physics of
Quantum Information", Springer, Berlin, 2000.
\bibitem{ic} M.A. Nielsen , I.L. Chuang, "Quantum Computation and Quantum
Information". Cambridge University Press, Cambridge, U.K. 2000.
\bibitem{na} N.A. Gershenfeld, I.L. Chuang, Bulk spin-resonance quantum computation, {\it Science} {\bf 275}, 350 (1997).
\bibitem{dg} D.G. Cory, A.F. Fahmy, T.F. Havel, Ensemble quantum computing by NMR spectroscopy, 
{\it Proc Natl Acad Sci. USA} {\bf 94} (1997) 1634.
\bibitem{kd} Kavita Dorai, Arvind, Anil Kumar, Implementing quantum-logic operations, 
 pseudopure states, and the Deutsch-Jozsa algorithm using noncommuting selective pulses in NMR, 
 {\it Phys Rev A.} {\bf 61}, (2000) 042306.
\bibitem{lo} T. S. Mahesh, Kavita Dorai, Arvind, Anil Kumar, Implementing Logic Gates 
 and the Deutsch-Jozsa Quantum Algorithm by Two-Dimensional NMR Using Spin- and Transition-Selective 
 Pulses, {\it J. Mag. Res.} {\bf 148}, 95 (2001).
\bibitem{ci} I.L. Chuang, N. Gershenfeld, M. Kubinec, Experimental implementation of fast quantum 
 searching, {\it Phys. Rev. Lett.} {\bf 80} (1998) 3408.
\bibitem{jo} J.A. Jones, M. Mosca, R.H.Hansen, Implementation of a quantum algorithm on 
a magnetic resonance quantum computer, {\it Nature.} {\bf 393} (1998) 344.
\bibitem{ap} L.M.K. Vanderspypen, M. Steffen, M.H. Sherwood, C.S. Yannoni,R. Cleve,
             and I.L. Chuang, Implementation of a three-quantum-bit search algorithm,
 {\it Applied Physics Lett.} {\bf 76}, 646 (2000).
\bibitem{qft} Y.S. Weinstein, M.A. Pravia, E.M. Fortunato, S.Llyod, and D.G. Cory,
 Implementation of the Quantum Fourier Transform, {\it Phys. Rev. Lett.} {\bf 86}, 1889 (2001).
\bibitem{nat} Lieven M.K. Vanderspyen, Matthias Steffen, Gregory Breyta, Costantino S.Yannoni,
 Mark H. Sherwood and Isaac L. Chuang, Experimental realization of Shor's quantum factoring 
 algorithm using nuclear magnetic resonance, {\it Nature.} {\bf 414}, 883 (2001).
\bibitem{kt} Kavita Dorai,T. S. Mahesh, Arvind, Anil Kumar, Quantum Computation using NMR, 
{\it Current Science.} {\bf 79}, 1447-1458 (2000).
\bibitem{ts} T. S. Mahesh, Anil Kumar, Ensemble quantum-information processing by NMR: 
Spatially averaged logical labelling technique for creating pseudopure states, 
 {\it Phys. Rev. A.} {\bf 64}, 012307 (2001).
\bibitem{ka} Kavita Dorai, Arvind, Anil Kumar, Implementation of a 
Deutsch-like quantum algorithm utilizing entanglement at the two-qubit level 
on an NMR quantum information processor, {\it Phys Rev A.} 63, 034101.
             (2001).
\bibitem{ran} Ranabir Das, T.S. Mahesh, and Anil Kumar, Efficient Quantum State Tomography for Quantum
Information Processing using  two-dimensional Fourier
Transform Technique, {\it Phys. Rev Lett.}(submitted).
\bibitem{ernst} Z.L. Madi, R. Bruschweiler and R.R. Ernst,
One- and two-dimensional ensemble quantum computing in spin Louivelle space,
{\it J. Chem. Phys.} {\bf 109}, 10603 (1998).
\bibitem{jm} J.A. Jones and E. Knill, Efficient Refocussing of One-Spin and Two-Spin Interactions 
 for NMR Quantum Computation, {\it J. Mag. Res.} {\bf 141}, 322(1999).
\bibitem{fu3} C.S. Yannoni, M.H. Sherwood, D.C. Miller, I.L. Chuang, L.M.K. Vandersypen,
               and M.G. Kubinec, Nuclear magnetic resonance quantum 
 computing using liquid crystal solvents, {\it Appl. Phys. Lett.} {\bf 75}, 3563 (1999).
\bibitem{fu1} A.K. Khitrin and B.M. Fung, Nuclear Magnetic Resonance 
Quantum Logic Gates using Quadruolar Nuclei, {\it J. Chem. Phys.} {\bf 112}, 6963(2000).
\bibitem{ne} Neeraj Sinha, T. S. Mahesh, K.V. Ramanathan, Anil Kumar,
              Toward quantum information processing by nuclear magnetic resonance:
 Pseudopure states and logical operations using selective pulses on an oriented spin
 3/2 nucleus, {\it J. Chem. Phys.} {\bf 114}, 4415 (2002).
\bibitem{mur} T.S. Mahesh, Neeraj Sinha, K.V.R.M. Murali, Malcom Levitt, K.V. Ramanathan
 and Anil Kumar, Quantum information processing by nuclear magnetic resonance: Experimental
implementation of half-adder and subtractor operations using an oriented spin-7/2 system,
{\it Phys. Rev. A} (accepted) (2002).
\bibitem{fu} B.M Fung, Use of pair of pseudopure states for NMR quantum computing,
{\it Phys. Rev. A} {\bf 63}, 022304 (2001).
\bibitem{ak}  T. S. Mahesh, Neeraj Sinha, K.V. Ramanathan, Anil Kumar,
               Ensemble quantum-information processing by NMR: Implementation
of gates and the creation of pseudopure states using dipolar coupled spins
as qubits, {\it Phys. Rev. A.} {\bf 65}, 022312 (2002).
\bibitem{mah} T.S. Mahesh, Neeraj Sinha, K.V. Ramanathan, and Anil Kumar, Quantum Information 
Processing by NMR: Manipulation of strongly coupled spins, {\it National Mag. Res. Symp.-8 Lucknow} (2002).
 
\bibitem{pou} D.G. Cory, A.E. Dunlop, T.F. Havel, S.S.Somaroo and W. Zhang, The effective Hamiltonian of 
Pound-Overhauser controlled-NOT gate, {\it quant-ph}/9809045 (1998).
\bibitem{abab} Simon C. Benjamin,Simple pulses for universal quantum computation with a Heisenberg ABAB chain,{\it
 Phys. Rev. A} {\bf 64}, 054303 (2001). 
\bibitem{lev} R. Freeman, T.A. Frenkiel and M.H. Levitt, Composite Z-pulses, {\it J. Mag. Res.} 
{\bf 44}, 409(1981).
\bibitem{ran} Ranabir Das, M.S. Thesis, IISc, 2002.
\bibitem{er} R.R. Ernst, G. Bodenhausen and A. Wokaun," Principles of Nuclear
        Magnetic Resonance in One and Two Dimensions", Clarendon Press,
          Oxford, U.K. 1987.
\bibitem{mod} Evan M. Fortunato, Marco A.Pravia, Nicolas Boulani, Grum Teklemariam,
Timothy F.Havel, and David G. Cory, Design of Strongly Modulating Pulses to Implememnt 
 Precise Effective Hamiltonian for Quantum Information Processing, quant-ph/0202065v1 (2002).

\end{thebibliography}
\end{document}